\documentclass[12pt]{iopart}
\usepackage{iopams}
\usepackage{graphicx}

\def\be{\begin{equation}}
\def\ee{\end{equation}}
\def\ba{\begin{eqnarray}}
\def\ea{\end{eqnarray}}
\def\lb{\label}
\def\nn{\nonumber}

\begin{document}

\title[Stochastic processes and complex Virasoro representations]{Stochastic processes with $Z_N$ symmetry and complex Virasoro
representations. The partition functions}

\author{Francisco C. Alcaraz$^1$, Pavel Pyatov$^2$ and Vladimir Rittenberg$^3$}\vspace{2mm}

\address {$^1$Universidade de S\~ao Paulo, Instituto de F\'{\i}sica de S\~ao
Carlos, Caixa Postal 369, 13560-590  S\~ao Carlos, S\~ao Paulo, Brazil}\vspace{-1mm}\ead{alcaraz@ifsc.usp.br}
\vspace{5mm}
\address{$^2$National Research University Higher School of Economics, 
Laboratory of Mathematical Physics,
20 Myasnitskaya street, Moscow 101000, Russia \&
Bogoliubov Laboratory of Theoretical Physics, Joint Institute
for Nuclear Research, 141980 Dubna, Moscow Region, Russia}\vspace{-1mm}\ead{pyatov@theor.jinr.ru}
\vspace{5mm}
\address {$^3$Physikalisches Institut, Universit\"at Bonn, Nussallee 12, 53115 Bonn, Germany}\vspace{-1mm} \ead{vladimir@th.physik.uni-bonn.de}

\begin{abstract}
In a previous Letter \cite{ARR} we have presented numerical evidence that a
Hamiltonian expressed in terms of the generators of the periodic
Temperley-Lieb algebra has, in the finite-size scaling limit, a spectrum
given by representations of the Virasoro algebra with complex highest
weights. This Hamiltonian defines a stochastic process with a $Z_N$
symmetry. We give here analytical expressions for the partition functions
for this system which confirm the numerics. For $N$ even, the Hamiltonian
has a symmetry which makes the spectrum doubly degenerate leading to two
independent stochastic processes. The existence of a complex spectrum
leads to an oscillating approach to the stationary state. This phenomenon is illustrated by an example.
\end{abstract}

\submitto{\JPA}
\maketitle

\phantom{a}
\vspace{10mm}

Considering $Z_N$ symmetric representations of the periodic Temperley-Lieb
algebra $PTL_L(x)$ \cite{L,MS,GL,MdSa}, we have defined a Hamiltonian as a linear combination of
the generators of this algebra. Taking $x = 1$, this Hamiltonian gives the time
evolution of a one-dimensional stochastic process. Looking at the finite-size
scaling spectra of this Hamiltonian, we have obtained numerical evidence for the
appearance of Virasoro representations with complex highest weights. Moreover, the
real part of the complex highest weights is smaller than the real highest
weights and hence dominate the large time behavior of the systems. This observation was a
big surprise and was the main content of a previous Letter \cite{ARR}. In the present
one, we give an analytic derivation of this result and present the partition
function for each sector of the model. We also present an application of our
results. For $N$ even we show that there is a symmetry in the model which makes the
spectrum for any lattice size to be doubly degenerate indicating the presence
of a zero fermionic mode.
This Letter is basically a continuation of the previous one \cite{ARR}, we did
nevertheless our best to make it self-consistent.
\bigskip

The $PTL_L(x)$ algebra has $L$ generators $e_k$,~ $k = 1,2,\dots , L$ satisfying the
relations:
\be
\lb{1}
e_k^2\, =\, x e_k, \qquad e_k e_{k\pm 1} e_k \, = \, e_k, \qquad [ e_k, e_\ell ]\, =\, 0, \;\; |k-\ell|>1,
\ee
with $e_{k+L} = e_k$. We take $L$ even only.
We consider two quotients of the algebra:
\be
\lb{2}
(AB)^N A\, =\, A,
\ee
where
\be
\lb{3}
A\, =\, \prod_{j=1}^{L/2} e_{2j}, \qquad B\, =\, \prod_{j=0}^{L/2-1} e_{2j+1},
\ee
and
\be
\lb{4}
A B A\,  =\, \alpha^2 A
\ee
with $\alpha =  e^{ i 2\pi r/N}$,~ $r = 0,1,\dots ,N-1$.
One can see that the quotient (\ref{4}) is a solution of eq.(\ref{2}) which defines
the first quotient. This observation will be crucial in obtaining the
partition functions mentioned above.
\bigskip

\begin{figure}[ht]
\phantom{a}\hspace{32mm}
\includegraphics[angle=0,width=0.3\textwidth] {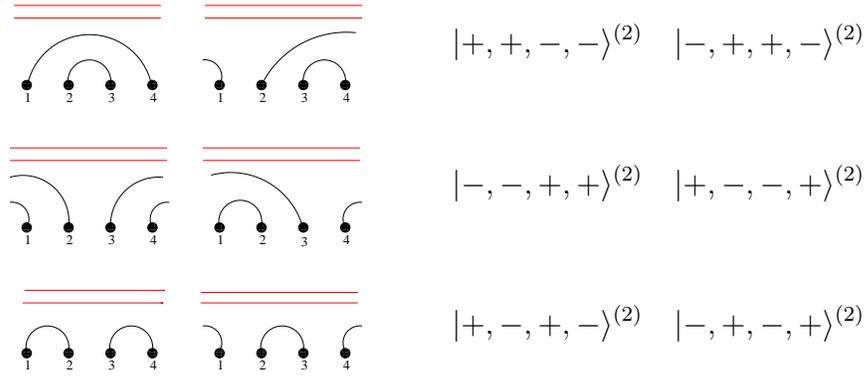}
\vspace{-5cm}
\ba
\nn
\hspace{67mm} |+,+,-,-\rangle^{(2)}\quad && |-,+,+,-\rangle^{(2)}
\\[12mm]
\nn
\hspace{67mm} |-,-,+,+\rangle^{(2)}\quad && |+,-,-,+\rangle^{(2)}
\\[12mm]
\nn
\hspace{67mm} |+,-,+,-\rangle^{(2)}\quad && |-,+,-,+\rangle^{(2)}
\ea
\vspace{-3mm}
\caption{
 The six link patterns configurations for $L = 4$ sites on a cylinder
and two circles without sites (noncontractible loops). The open arches and
circles meet behind the cylinder. The corresponding spin presentations of the same
link patterns are given on the right.}
\label{fig1}
\end{figure}

In order to get the $Z_N$ symmetric representations of (\ref{1}) and (\ref{2}), we consider
$N$ copies of a one-dimensional periodic system with $L$ sites.
Each copy consists of $L\choose L/2$ configurations of link patterns on a cylinder and  $n$  noncontractible loops ($n = 0,1,\dots , N - 1$) on the same cylinder. This is the vector space in which the generators of the
$PTL_L(x)$ algebra act. It has the dimension $N\times {L\choose L/2}$. In Fig.\ref{fig1} we show the $6$
configurations for $L = 4$ and $n = 2$.

An alternative way to label the states in the vector space is to use the spin
representation in which the slopes in the arches are used $+(-)$ for the beginning
(ending) of an arch. The number of non-contractible loops is indicated by a
supplementary label. This notation is also given in Fig.\ref{fig1}.

\begin{figure}[h]
\centering
\includegraphics[angle=0,width=0.33\textwidth] {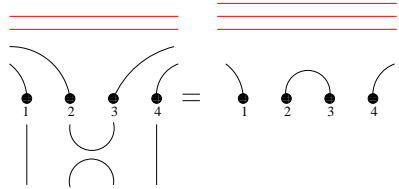}
\caption{
The action of the $e_2$ generator acting the bond between the sites 2
and 3 which are the end of an arch of the size of the system $L = 4$. A new
circle is created on the cylinder and one moves from the copy $n = 2$ to the
copy $n = 3$.}
\label{fig2}
\end{figure}

The action of the generators $e_k$ on the link patterns for a given copy $n$ is the
same as the one used for the usual (non-periodic) Temperley-Lieb algebra \cite{M} with
one exception. If the generator acts on the bond connecting the beginning and the
end of an arch having the size of the system $L$, one obtains a configuration of the
copy $n + 1$ (see Fig.\ref{fig2}).

In order to obtain the $Z_N$ representation of the $PTL_L(x)$ algebra with the
quotient (\ref{2}) one identifies the copy $n = N$ with the copy $n = 0$.
\bigskip

In what follows, we take the parameter $x$ in the $PTL_L(x)$ algebra equal to
one. With this choice the Hamiltonian
\be
\lb{5}
H\, =\, \sum_{k=1}^L (1-e_k)
\ee
gives the time evolution of a stochastic process. We want to stress that the
properties of the spectra which are going to be discussed below, stay valid for any
value of $x$.

Notice that for $N$ even the $Z_N$ symmetric representation decomposes into a pair of identical
$\frac{N}{2} \times {L\choose L/2}$ -dimensional irreps.\footnote{For $N$ odd the $Z_N$ representation is irreducible.}
To see this, consider two linear transformations  on the $Z_N$ space.

The  transformation $X_1$ acts diagonally in a following way. We start considering configurations with no arches hidden 
in the back of the cylinder (the first and fifth link patterns in Fig.\ref{fig1}).
These configurations get a factor of $(-1)^n$. The configurations translated with one lattice unit
get a factor $(-1)^{(n+1)}$ (the second and sixth configurations in Fig.\ref{fig1}). The next translated configurations
get again the factor $(-1)^n$ and so on and so forth.


Transformation $X_2$ permutes copies as follows:
$$
|\dots\rangle^{(2k)}\leftrightarrow  |\dots\rangle^{(2k+1)}
,\qquad  k=0,1,\dots , \frac{N}{2}-1.
$$

The two transformations constitute the algebra:
\be
\lb{8}
(X_1)^2\, =\, (X_2)^2\, =\, \mbox{id},\qquad X_1 X_2 \, +\, X_2 X_1\, =\, 0.
\ee
This algebra has only 2-dimensional equivalent irreducible representations. Since $X_1$ and
$X_2$ commute with the action of the $PTL_L(x)$ on the $Z_N$ symmetric link patterns, it follows that for $N$ even, the spectrum of $H$ is doubly degenerate.
To illustrate this observation, let us take $L = 4$
and $N = 2$. The Hamiltonian splits into two stochastic Hamiltonians having each
six states. The first one having the states $|++--\rangle^{(0)}$, $|--++\rangle^{(0)}$, $|+-+-\rangle^{(0)}$,
$|-++-\rangle^{(1)}$, $|+--+\rangle^{(1)}$ and $|-+-+\rangle^{(1)}$, the second one having the six states in which
the copies $0$ and $1$ are permuted.
\bigskip

We are interested in the spectra of $H$ in the finite-size scaling limit.
Let us keep in mind that in a stochastic process, the energies coincide with the energy
gaps since the ground-state energy is zero for any system size.
Since $H$ is
invariant under translations ($e_k \mapsto e_{k+1(\mbox{\tiny mod} L)}$) and the cyclic rotations $Z_N$
($|\dots\rangle^{(n)}\mapsto |\dots\rangle^{(n+1\mbox{\tiny (mod $N$)})}$),
one has ${N\times L}$ sectors
labeled by $p = 0, \pm 1,\pm 2,\dots$ corresponding to the momenta $P = 2\pi p/L$ and by
$r = 0,1,\dots , N - 1$, labeling the irreps of $Z_N$. If $E_p^r(q)$,~ $q = 1,2,\dots ,$ are
the energy levels in the sector $(p,r)$, the scaling dimensions $x_p^r(q)$ are
given by $\lim_{L\rightarrow \infty}(E_p^r(q) L) = 2\pi v_s x_p^r(q)$, with the sound velocity
$v_s = 3\sqrt{3}/2$.

In \cite{ARR} we have diagonalized numerically $H$ for $N = 3$. We went up to $L = 30$
and looked at the lowest excitations. In the $r = 0$ sector we confirmed the
expected value $x_0^0 (1) = 0.25$. The surprise came when we looked at the $r = 1$
sector where we found:
\be
\lb{9}
x_0^1(1) = 0.03905 + 0.08753 i, \qquad x_0^1(2) = 0.14908 -0.11806 i,
\ee
i.e. complex values.
For $N$ even we found $E_p^r(q) = E_{p+L/2}^{r+N/2}(q)$ which is a consequence of
the symmetry (\ref{8}).
\bigskip

We present now our new results. In order to obtain the partition function in each
sector $r$, we use the fact that the $Z_N$ representation of the algebra with the
quotient (\ref{2}) can be decomposed into $N$ representations of the quotient (\ref{4}) \cite{MS}.
The representations of the quotient (\ref{4}), in the link patterns vector space, are
obtained by considering a single copy but changing the action of the generators
when they act on a bond connecting the beginning and the end of an arch of the
system size $L$ like in Fig.\ref{fig2}. Instead of adding a non-contractible loop, one
multiplies the state in the right hand side of the figure by a fugacity
$\alpha = \exp(2\pi i r/N)$. It was shown \cite{L} that the quotient (\ref{4}) admits also a
representation in the standard spin $1/2$ basis (not to be confused with the one used in Fig.\ref{fig1}) and the Hamiltonian (\ref{5}) can be written in
this basis. By performing a similarity transformation, the Hamiltonian is the
XXZ quantum chain with a twist:
\ba
\nn
\phantom{a}\hspace{-22mm}
e_k &=& \sigma^+_k \sigma^-_{k+1}\, +\, \sigma^-_k \sigma^+_{k+1}\, +\,
\frac{1}{4}(1-\sigma^z_k \sigma^z_{k+1})\, +\, i \frac{\sqrt{3}}{4}(\sigma^z_{k+1}-\sigma^z_k),
\;\; k=1,2,\dots ,L-1,
\\
\lb{10}
\phantom{a}\hspace{-22mm}
e_L &=& e^{ i 2\pi\phi}\sigma^+_L \sigma^-_1\, +\, e^{ -i 2\pi\phi}\sigma^-_L \sigma^+_1\, +\,
\frac{1}{4}(1-\sigma^z_L \sigma^z_1)\, +\, i \frac{\sqrt{3}}{4}(\sigma^z_1-\sigma^z_L).
\ea
The twist $\phi$ is related to the parameter $\alpha$ by the relation:
\be
\lb{11}
\alpha\, =\, 2\cos(\pi \phi).
\ee

The vector space of the ${L \choose L/2}$ link patterns configurations corresponds to the
$S^z = \sum_{k = 1}^L \sigma_k^z = 0$ sector of the spin vector space.

The Hamiltonian (\ref{5}),(\ref{10}) is  integrable using the Bethe Ansatz and the scaling
dimensions (highest weights of Virasoro representations) are known \cite{ABB}. 
They are given by the Gaussian model. In the $S^z = 0$ sector they are:
\ba
\lb{12}
&&x = \frac{3}{4} (s+\phi)^2 - \frac{1}{12} + m + m', \quad p=m-m',
\ea
where
$s,m,m'=0,\pm 1,\pm 2,\dots$.

From (\ref{11}) we see that for N even $\alpha = e^{i2\pi r/N}$ and
$\alpha=e^{i2\pi(r +N/2)/N} = - e^{i2\pi r/N}$
give the same value for the twist $\phi$ and therefore
the spectrum of the Hamiltonian (\ref{5}) is doubly degenerate, in agreement with our
previous observation. Moreover, for the sectors $r$ not equal to $0$ or $N/2$ the
values of $\phi$ obtained from (\ref{11}) are complex, henceforth the scaling
dimensions (critical exponents) (\ref{12}) are complex too.

As a check, taking $N = 3$ and $r = 1$, from (\ref{11}) one gets:
$$
\phi = - 0.426642 - 0.137279 i
$$
from which we get using (\ref{12}) with $s = 0$ and $1$
\be
\lb{13}
\phantom{a}\hspace{-1cm}
x_0^1(1) = 0.03990499 + 0.087853992 i ; \quad x_0^1(2) = 0.1490874 - 0.11808136 i
\ee
in excellent agreement with the values (\ref{9}).
\bigskip

We have to stress that although the spectrum of the $Z_N$ symmetric Hamiltonian
splits into $N$ sectors, the stochastic process doesn't. The condition of
positivity of the wave function describing the probability distribution
function, mixes the sectors. 
\bigskip

\begin{figure}[t]
\vspace{16mm}
\centering
\includegraphics[angle=0,width=0.7\textwidth] {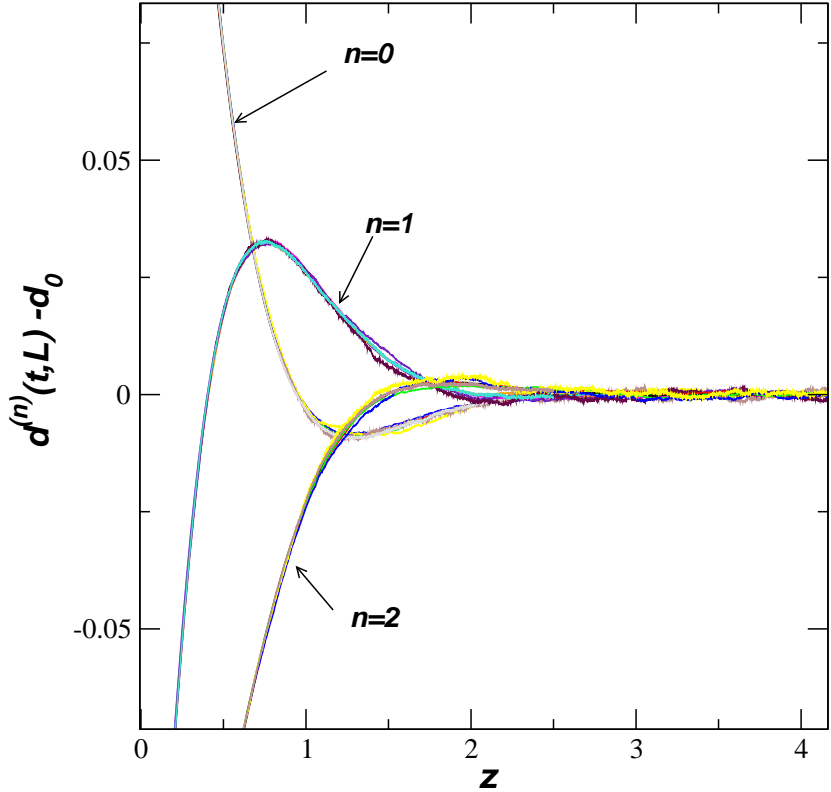}
\caption{
The density of "peaks" $d^{(n)}(z)$ in the copy $n$ as a function of
$z =2\pi v_s t/L$ for $N = 3$, and various lattice sizes $L = 40, 80, 160, 500, 1000,
2000$ and $4000$. $d_0$, the value of the density in the stationary state, is
subtracted.}
\label{fig3}
\end{figure}

The existence of complex scaling dimensions has consequences on the time
behavior of various correlators showing oscillatory phenomena, more so since
their real part is smaller than real scaling dimensions of the $r = 0$ sector. To
illustrate the phenomenon, we looked at the density of "peaks" $d^{(n)}(t,L)$ in
different copies. Those are $+-$ pairs which in the Dyck paths picture of
the link patterns \cite{AR} correspond to peaks in the paths. 
This local observable is measured easily in Monte Carlo simulations. The time dependence of the
"peaks" in various sectors is determined by the initial conditions. For large values
of $t$ and large lattice sizes we expect $d^{(n)}(t,L)$ to be a function of $t/L$:
\be
\lb{14}
d^{(n)}(t,L)\, =\, d_0\, +\, \sum_k A_k \cos{[\mbox{Im}(x_0^{(n)}(k)) z]} e^{-\mbox{\small Re}(x_0^{(n)}(k)) z},
\;\; z=\frac{2\pi v_s t}{L},
\ee
where $d_0$ is the density of "peaks" in the stationary state which is the same
for each copy $n$ and the $A_k$'s are dependent on the initial conditions. We have
computed $d^{(n)}(t,L) - d_0$ using Monte Carlo simulations in the case $N = 3$ for
different lattice sizes. The initial state was the configuration $|+,-,+,-,\dots ,+,-\rangle^{(0)}$ in
the copy $n = 0$. The results are shown in Fig.\ref{fig3}. One can see that, as expected, the
densities are dependent on $z$ only.

\begin{figure}[t]
\vspace{16mm}
\centering
\includegraphics[angle=0,width=0.7\textwidth] {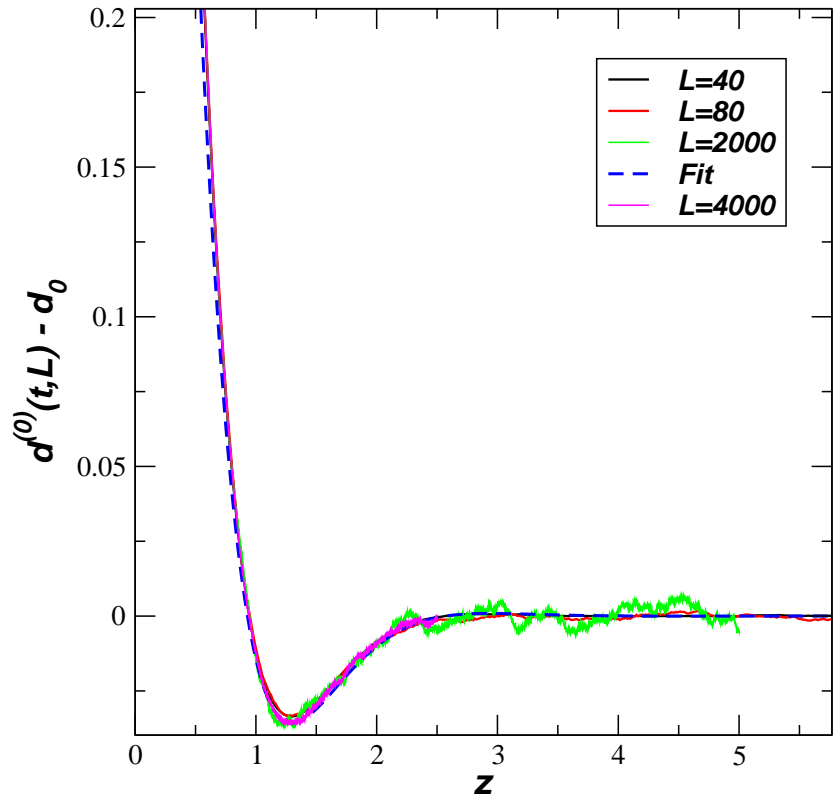}
\caption{
A fit to $d^{(0)}(z) - d_0$ as a function of $z$, using the parameterization (\ref{15}) and (\ref{16}). Monte
Carlo data for lattice sizes $40,80,2000$ and $4000$ were used. The fit is
indicated by dashed line.}
\label{fig4}
\end{figure}

Encouraged by this observation, we did a fit
to the $n = 0$ data (see Fig.\ref{fig4}) using the parameterization
\be
\lb{15}
d^{(0)}(z)\, -\, d_0\, =\, e^{-a z}\frac{\cos{b(z-z_0)}}{\cos{b z_0}}
\ee
and obtain:
\be
\lb{16}
a = 0.1379\, ,\quad  b = 0.1107\, ,\quad z_0 = 0.060\, .
\ee

These values are compatible with $x_0^1(2)$ given in (\ref{13}). To our knowledge, it is for the
first time that expressions like (\ref{15}) appear in a conformal invariant theory.
\bigskip

Before closing this Letter, we would like to notice that in a seminal paper 
Saleur and Sornette \cite{LL} have suggested the possible existence of 
complex critical exponents in non unitary conformal field theories. We have shown that they indeed exist.

\section*{Acknowledgements}

This work was supported in part by the joint DFG and RFBR grants no.RI 317/16-1 and no.12-02-5133-NNIO, by FAPESP and CNPq (Brazilian Agencies), and a grant of the Heisenberg-Landau program. PP was also supported by the RFBR grant 14-01-00474 and  by the  Higher School of Economics Academic Fund grant 14-09-0175. FCA thanks the Bethe Center of the Bonn University for partial financial support.

\section*{References}



\begin{thebibliography}{WW}
\label{refer}


\bibitem{ARR} Alcaraz F C,  Ram A and Rittenberg V 2014
{\it J. Phys. A} {\bf 47} 212003

\bibitem{L} Levy D 1991 {\it Phys. Rev. Lett.} {\bf 67} 1971


\bibitem{MS} Martin P P and Saleur H 1993 {\it Comm. Math. Phys.} {\bf 158} 155;
1994 {\it Lett. Math. Phys.} {\bf 30} 189

\bibitem{GL} Graham J J and Lehrer G I 1998 {\it  L'Enseignment Math.} {\bf 44} 173; 2002
{\it Compositio Math.} {\bf 133} 173; 2003 {\it Ann. Sci. Ecole Norm. Sup.} {\bf 36} 479

\bibitem{MdSa} Morin-Duchesne A and Saint-Aubin Y 2013 {\it J. Phys. A} {\bf 46} 285207

\bibitem{M} Martin P 1990 {\it Potts models and related problems in statistical mechanics} World Scientific

\bibitem{ABB} Alcaraz F C, Barber M N and Batchelor M T 1989 {\it Ann. Phys.} {\bf 182} 280

\bibitem{AR} Alcaraz F C and Rittenberg V 2013 {\it J. Stat. Mech.} P09010

\bibitem{LL} Saleur H and Sornette D 1996 {\it J. Phys. I} {\bf 6} 327

\end{thebibliography}
\end{document}